\documentclass[10pt,prd,twocolumn,showpacs,amsmath,amssymb,nofootinbib,preprintnumbers]{revtex4-1}
\usepackage{wasysym}
\usepackage{graphicx}
\usepackage{dcolumn}
\usepackage{bm}
\usepackage{amssymb}
\usepackage{epsfig}
\usepackage{slashed}
\usepackage{graphicx,amsmath,amssymb,bm}
\newcommand{\be}{\begin{equation}}
\newcommand{\ee}{\end{equation}}
\newcommand\beq{\begin{eqnarray}}
\newcommand\eeq{\end{eqnarray}}
\newcommand\eqn[1]{\label{eq:#1}}
\newcommand\eq[1]{eq. (\ref{eq:#1})}

\begin{document}
 \title{Matrix Flavor Brane and Dual Wilson Line}
\author{Andreas Karch}
\email[Email: ]{akarch@uw.edu}

\affiliation{Department of Physics, University of Washington, Seattle, WA
98195-1560, USA}

\author{Sichun Sun}
\email[Email: ]{sichun@uw.edu}
\affiliation{Institute for Nuclear Theory, Box 351550, Seattle, WA 98195-1550, USA}

\preprint{INT-PUB-13-047}

\begin{abstract}
We study a novel non-Abelian matrix configuration of probe D-branes in AdS$_5$. This configuration gives rise to a new D-brane phenomenon related to the known ``Myers effect" in the context of holography. It is dual to a deformation of the field theory by a Wilson line threaded fermion bilinear. We study the two point function of these short Wilson lines from both the non-Abelian DBI action and a classical string world sheet calculation and identify the region where they agree. We also study a related configuration where the non-Abelian nature of the embedding functions is enhanced by a background flux as in the Myers effect.
\end{abstract}

\date\today

\maketitle

\section{Introduction}

In string theory a pointlike object (a D0 brane) can be polarized into a membrane with a 2+1 dimensional worldvolume by an external ``electric" field coupling to membrane charge. The resulting spherical membrane has zero net membrane charge, but carries the original D0 brane charge as an induced charge density on its worldvolume. This effect is usual referred to as the Myers-effect \cite{Myers:1999ps}. From the D0 point of view, the membrane appears as a non-Abelian configuration for the matrices representing the coordinates of the D0 branes. The Myers effect can be straightforwardly generalized to higher dimensional D-brane defects. In particular, a D$p$ brane, that is a defect with a $p+1$ dimensional worldvolume, can be polarized into a D$(p+2)$ brane by field strength coupling to D$(p+2)$ brane charge. Again, this brane polarization crucially relies on the non-Abelian nature of the underlying D-brane action.

In this work we show that, in the context of holography \cite{Maldacena:1997re,Gubser:1998bc,Witten:1998qj}, this same brane polarization can be forced upon a D-brane, in the absence of any fluxes, by turning on certain sources in the boundary field theory. We study $N_f$ D$p$-brane probes\footnote{In the limit of weak string coupling $g_s$, the D-brane has a negligible tension (the tension grows as $1/g_s$, but Newton's constant goes to zero as $g_s^2$) and so the probe does not backreact on the background geometry.} embedded in an asymptotically anti-de Sitter (AdS) space. Via holography the probe branes in AdS can be mapped to a large $N_c$ conformal field theory coupled to $N_f$ fundamental flavor fields \cite{Karch:2000gx,Karch:2002sh}.

In general, the fundamental flavor fields can either occupy all the field theory directions or be localized on a lower dimensional defect. For our work however it is crucial that the flavor branes realize a defect field theory. This allows us to separate the defects along the field theory directions. For $N_f$ coincident flavor branes, the field theory has a $U(N_f)$ global symmetry rotating the localized flavors into each other. There are $N_f^2$ different mass terms one can turn on for the $N_f$ flavors. In the holographic dual a single mass term  corresponds to geometric deformations of the flavor brane; more concretely it maps to a ``slipping mode" that describes how the probe D-brane in the bulk is embedded in the internal space, the $S^5$ of AdS$_5$ $\times$ $S^5$. Separating the defects breaks this global symmetry to a $U(1)^{N_f}$. In the bulk, we now have $N_f$ well separated probe branes. Only the $N_f$ diagonal mass terms can be realized as geometric deformations of these probe branes. They correspond to adding a fermion bilinear to the field theory action made out of fermions localized on a single defect. What we will focus on in this work is adding bilinears made from fermions localized on different defects. By gauge invariance, such non-local bilinears need to be threaded by a Wilson-line, as is well known from studies of the Sakai-Sugimoto model \cite{Sakai:2004cn}. When these non-local mass terms are turned on, the dual probe brane system is forced into a non-Abelian configuration. A detailed quantitative study of these configurations is difficult due to the poorly understood nature of the non-Abelian action governing coincident D-branes. We are however able to reliably study the asymptotic behavior of the brane embedding. This allows us to uncover a novel UV behavior of the embedding functions due to the non-local nature of the deformation. We also are able to calculate the short-distance behavior of the 2-pt function of the Wilson-line-threaded fermion bilinears. This can be compared with a calculation of the same objects using Maldacena's description of Wilson lines via string worldsheets \cite{Maldacena:1998im} and perfect agreement is found.

Several examples of non-Abelian D-brane configurations in AdS have previously been documented in the literature. Maybe the most famous application of non-Abelian configurations in the context of holography is the realization of confinement via the Polchinski-Strassler solution \cite{Polchinski:2001tt} - here the background D3 branes themselves puff into 5-branes. The work most closely related to what we attempt to do here is probably the recent paper \cite{Kristjansen:2012ny}. The authors presented a D7 brane embedding with induced D5 brane charge which asymptotically (that is close to the AdS boundary) wraps a vanishing 2-cycle. As a consequence the dual field theory contains a defect with the matter content characteristic of a D5 brane probe, not a D7 brane. The puffing into a D7 deep in the interior of AdS corresponds to a novel state in the defect theory described by the D5 brane. While both of these papers, and others similar to them, can be thought of as holographic implementations of the Myers effect, they always allowed a description in terms of a higher dimensional probe brane with flux. Instead of describing a non-Abelian configuration of D$p$-branes, the systems have an alternative description in terms of D$(p+2)$ branes with a D$p$-brane charge inducing flux. This higher-dimensional action was effectively Abelian. Even though in principle coincident probe branes are described by a non-commutative action the configurations studied did not require any genuinely non-commuting matrices when viewed from the higher dimensional perspective.
As far as we can tell, the configurations we study in this paper require a genuine non-Abelian D-brane configuration in AdS. The non-Abelian nature is mandatory given the couplings we turn on in the boundary field theory. Holographic probe branes with non-Abelian gauge field configurations have been studied before e.g. in \cite{Erdmenger:2007ja,Myers:2008me,Ammon:2009fe}, but we believe our system is the cleanest example of a probe brane with genuinely non-Abelian scalars and hence a non-geometric embedding.

We will analyze two quite distinct setups which realize this general idea. In the first part of the paper we study a supersymmetric defect field theory with D5 brane probes with 4 ``ND" directions in a D3 brane background. The defect field theory dual to this setup describes 2+1 dimensional hypermultiplets coupled to 3+1 dimensional ${\cal N}=4$ super Yang-Mills theory. We establish the duality between the non-local deformations of the field theory and the non-Abelian D-brane configurations. We also analyze the short-distance behavior of correlation functions in this theory and compare to a classic Wilson line calculation.

In the second part of the paper we want to analyze how the non-Abelian nature of the brane embeddings is enhanced when we turn on additional background fluxes as are present in the original Myers effect. For this purpose we analyze D5 probe branes with 6 ``ND" directions. In this case the defect localized matter is purely fermionic: 2 Dirac fermions localized on 1+1 dimensional defects in the 3+1 ${\cal N}=4$theory. Turning on an explicit 9-form field strength in the bulk helps to polarize the D5 probe into a D7 probe brane, just as in the original Myers setup. The background flux allows us to get non-trivial commutators form the Wess-Zumino (WZ) term in the action already at cubic order in the fields, whereas commutators only entered at quartic order in the supersymmetric setup. To this order only the Abelian terms contribute to the Dirac-Born-Infeld (DBI) part of the action. We once more can reliably analyze the UV asymptotic behavior of the solution using a small field expansion of the action.
The separation of the defects allows to lift some of the instabilities present in the non-supersymmetric 6 ND system above the Breitenlohner-Freedman bound. For this stabilization to work with a small brane separation we require the 9-form flux to be significantly different from zero. Correspondingly, we need to construct the probe brane embeddings in a gravitational background where the full backreaction of the 9-form flux has been taken into account. Fortunately, such a gravitational background has been constructed recently \cite{Azeyanagi:2009pr,Mateos:2011ix}. This background is dual 4D super Yang-Mills (SYM) theory with a spatially varying theta angle.  In this setup we can study the non-local mass term reliably in the limit of small brane separation. While this is sufficient to stabilize the off-diagonal slipping modes, the diagonal slipping modes won't get stabilized by our construction; more ingredients are needed for a fully stabilized solution, e.g. we could turn on internal fluxes \cite{Myers:2008me,Bergman:2010gm}. Also, while the inclusion of the flux allows us to analyze stabilization of the off-diagonal modes in the UV reliably, we still need to resort to a non-linear action in the bulk in order to find a complete brane embedding. In order to numerically find a full solution for the brane embedding in this case, we make the ad-hoc assumption that commutators coming from the DBI part of the action are negligible and only the commutators from the WZ term coupling to the external flux are kept.

The paper is organized as follows: in the next section we introduce our main example: the supersymmetric D3/D5 system together with its non-Abelian deformation. In section III we calculate the 2-pt function in this system and compare to a classic Wilson line computation. In section IV we study non-supersymmetric branes in the background of polarizing fluxes. We conclude in section V.

\section{Non-Abelian supersymmetric D3/D5 system without flux}

Let us first review the Abelian solution for $N_f$ D5-branes intersecting $N_c$ D3-branes in $2+1$ dimensions as first introduced in \cite{Karch:2000gx,DeWolfe:2001pq}. The directions occupied by D3 and D5 brane are displayed in the following table.

\begin{eqnarray}
\begin{array}{c|rrrrrrrrrr|}
  {}  & 0  &1 &2 &3 &4& 5  & 6 & 7 & 8& 9 \\ \hline
 D3 & \times& \times& \times& \times&&&&&&\\
  D5 & \times& \times&\times&& \times&\times& \times&& &\\\hline
\hline
\end{array}\end{eqnarray}

The gravity description will be studied in the probe limit, with $N_f\ll N_c$. Then for large $N_c$ and $g_s N_c$, we can treat D5-brane as probes in the asymptotically AdS$_5$ geometry produced by large number of D3-branes

Let us consider the AdS/CFT dual of our D3-D5 brane system with a single D5 brane probe to warm up. The background geometry is the asymptotically flat, zero temperature D3-brane solution:
\beq
\label{d3metric}
ds^2=f^{-\frac{1}{2}}(-dt^2+d\vec{x}^2)+f^{\frac{1}{2}}(dr^2+r^2d\Omega_5^2)
\eqn{G}\eeq
We can rewrite $\Omega_5$ above as :
\beq
dr^2+r^2 d\Omega_5^2=d^2\rho+\rho^2 d\Omega_2^2+dx_7^2+d^2x_8+d^2x_9
\eeq
with
\beq
\eqn{fis} f=\frac{R^4}{(\rho^2+x_7^2+x_8^2+x_9^2)^2} = \frac{R^4}{r^4}.
\eeq
In what follows we will usually work in units where $R=1$.
The D5 probe brane spans $x_0$, $x_1$ and wraps the $S^3$ above and lives on a curve $x_9(\rho)$. For the simplest solution we set $x_7=x_8=x_3=0$. The pull-back metric on the D5 brane worldvolume is:
\beq
\nonumber && ds_6^2=\frac{(-dt^2+dx_1^2+dx_2^2)}{f^{1/2}}\\ &&
+f^{\frac{1}{2}} \left [ (1+x_9'^2)d\rho^2 + \rho^2 d\Omega_2^2 \right ]
 \eeq
Without worldvolume gauge field, the action describing the embedding of the branes, the Dirac-Born-Infeld (DBI) action, is given by:
\begin{eqnarray}
S_{D5}&=&-N_fT_5\int d^6\xi \sqrt{-g_6}
\\ \nonumber &=&-N_fT_5 Vol(S^2)\int d^2x d\rho \rho^2\sqrt{1+x_9'^2}
\label{d5action}
\end{eqnarray}
From the linearized equations of motion for $x_9$
\beq
 2\rho x_9'(\rho)+\rho^2 x_9''(\rho)=0
\eeq
one can read off its asymptotic behavior:
\beq
x_9(\rho)=C_1 +C_2\rho^{-1}.
\eeq
This is the right fall-off to correspond to conformal dimension $\Delta=1$ or $\Delta=2$, depending on the quantization scheme. For the supersymmetric D5 brane system one can see \cite{DeWolfe:2001pq} that supersymmetry requires the correct dimension to be $\Delta=2$. Another way to arrive at this result is to study the angular description of this slipping mode in terms of $\theta$ in $x_9=r \, sin\theta$, which obeys the equations of motion of a scalar with $m^2=-2$ in AdS$_4$. This fluctuation is dual to the fermion mass term on the defect.

\subsection{Multi-brane solution}
For $N_f$ coincident D-branes the action \eqref{d5action} has to be generalized to the non-Abelian DBI action. In this section we'll closely follow the discussion in  reference \cite{Myers:1999ps}.
In terms of $N_f$ by $N_f$ matrix valued coordinates and gauge fields one has:
\begin{eqnarray}
S_{D5}&=&-T_5 \int d^6x  \label{action}\\  \nonumber && Tr \left (\sqrt{-det(P[G_{ab}])det(Q^i_j)} \right )
\end{eqnarray}
where
\beq
& Q^i_j\equiv \delta^i_j+\frac{i}{\lambda}[x^i,x^k]G_{kj}.\\
& \lambda=2\pi \ell^2_s
\eqn{fac}\eeq
In principle we have to commit to a procedure of how to deal with the ordering ambiguities of the non-Abelian action. As long as  the fields remain small, we can expand the action in a power series in the fields.
The first time a commutator arises from the DBI action is at quartic order in the fluctuating fields. Up to this order the action is free of ordering ambiguities. So as long as our fields are small, we have full control over the non-Abelian action. This will be sufficient for us to analyze the UV behavior of the brane fluctuations as well as for the computation of the short distance behavior of the two-point function.
Expanding the non-Abelian structure to the next leading order in the embedding scalars one has for diagonal  $G_{ab}$:
\beq
\label{qis}
\sqrt{det Q^i_j}=1-\frac{1}{4\lambda^2}[x^i,x^j][x^i,x^j] G_{ii}G_{jj}+...
\eeq
To see the non-Abelian effect, we write down non-Abelian action for D5 brane as:
\begin{eqnarray}
S_{D5}&=&-N_fT_5\int d^6\xi \sqrt{-g_6}
\\ \nonumber &=&-N_fT_5 4 \pi \int d^2x d\rho \rho^2\sqrt{1+x_9'^2+x_8'^2 +f^{-1} x_3'^2}\\
&& (1-\frac{1}{4\lambda^2}[x^i,x^j][x^i,x^j] G_{ii}G_{jj}) \nonumber \\ && \quad i,j= 3,8,9
\end{eqnarray}
From now on we specialize to the case of $N_f=2$. We wish to describe a configuration in which the two defects are separated in the field theory directions. The field theory coordinate of the two branes is given by the constant asymptotic value of the fluctuating field $x_3$ in the bulk. In order to have a geometric interpretation of the defect positions it is convenient to use an $SU(2)$ flavor rotation diagonalizing $x_3$. We can chose the origin in the $x_3$ direction to be in the middle between the two defects, which we take to be separated by a distance $2d$. With this we asymptotically have
\begin{equation}
x_3 \sim d \sigma_3.
\end{equation}
$x_7$, $x_8$ and $x_9$ encode the triplet of mass terms one can add for a hypermultiplet. For simplicity, we only consider the case $x_7=0$. The asymptotic behavior of $x_8$ and $x_9$ now encode real and imaginary part of the standard complex mass terms we add for the flavors on the defects. If the leading behavior for $x_8$ and $x_9$ is diagonal, we added only local mass terms on the two defects. However, if we chose to add non-local Wilson-line threaded mass terms, they are off-diagonal. They can not be diagonalized without ruining the geometric interpretation of $x_3$. The addition of these mass terms forces us to find a genuinely non-Abelian solution.

For concreteness, let us consider the following $SU(2)$ ansatz for a non-commuting solution, separating the radial parts and matrix structure as $x_3=X_{1}(\rho)\sigma_3$, $x_8=X_{2}(\rho)\sigma_1$, $x_9=X_{3}(\rho)\sigma_2$, with $[\sigma_i,\sigma_j]=2i \epsilon_{ijk}\sigma_k$ and $\sigma_{i}^{2}=1$. We insist that $X_2=X_3$, that is we chose our off-diagonal mass to be proportional to $\sigma_1+\sigma_2$ and assume that any condensate that forms is of the same form. Note that with this choice the radial functions are symmetric under $x_8\leftrightarrow x_9$. With this ansatz we can evaluate the commutators as
\beq
[x_i,x_j]=2i \epsilon_{ijk}\sigma_k X_iX_j .
\eeq
Substituting this ansatz into equation of motion, one arrives at:
\begin{eqnarray}
\nonumber &S_{D5}  = -N_fT_5 4 \pi \int d^2x d\rho \rho^2\sqrt{1+X_2'^2+X_3'^2 +f^{-1} X_1^2} \\
\nonumber & \left[ 1+\frac{2}{\lambda^2}(X_1^2X_2^2+X_1^2X_3^2+X_2^2X_3^2 f)\right]
\end{eqnarray}
\\
For the radial functions we now take an ansatz
\beq
\nonumber &X_{1}(\rho)\rightarrow d+\epsilon g(\rho)\\
 & X_{2,3}(\rho)\rightarrow\epsilon k(\rho)
\eeq
so that $g$ vanishes near the boundary.
To determine the near boundary UV behavior of the solution we can linearize the equations of motion in $\epsilon$:
\beq
\nonumber &-4\frac{d^2}{\lambda^2} \rho^2 k(\rho)+ 2\rho k'(\rho)+\rho^2 k''(\rho)=0\\
&6g'(\rho)+\rho g''(\rho)=0
\eeq
which yields solutions:
\beq
\label{mresult}
\nonumber &g(\rho)=\frac{C_1}{5\rho^5}+C_2\\
\nonumber &k(\rho)=C'_1 e^{-m \rho }/ \rho+C'_2  e^{m \rho }/ \rho\\
&\text{with} \quad m=(2d)/\lambda
\eeq
where $m$ has dimension of energy. Since $\lambda^{-1}$ is the tension of a fundamental string,
$m$ is exactly the mass of a piece of string stretched between the two defects. This behavior of $x_8$ and $x_9$ in the UV may look strange at first sight, but we will give a possible interpretation of the exponential factors in the next section. As a consistency check one should note that if one takes the limit where $m\rightarrow0$, which corresponds to an Abelian solution with two branes on top of each other, one recovers the result from the last section.

\subsection{Field theory description}

The field theory dual to this probe brane setup is now easy to describe. The D3 branes themselves give the usual glue theory: ${\cal N}=4$ SYM in the limit of a large number of colors $N_c$ and large 't Hooft coupling. Each flavor brane gives rise to a single 2+1 dimensional Dirac fermion in the fundamental representation of $SU(N_c)$ and their superpartners localized along a sheet-defect inside the 3+1 dimensional SYM.

The $N_f$ fermions do not all have to be localized on the same defect. In our construction, we separated the defects along the $x_3$ direction. From the holographic bulk point of view $x_3$  is a fluctuating scalar degree of freedom living on the worldvolume of the probe and the asymptotic value $x_3$ takes for the $i$-th brane maps to the position of the $i$-th defect in the field theory.

In the non-Abelian setting of $N_f=2$ flavor branes, in the bulk all coordinates are 2 by 2 matrix. But we can always chose a gauge in the bulk in which one of them is diagonal. As $x_3$ has a geometric interpretation in the field theory, we chose $x_3$ to be diagonal; the two entries then correspond to the field theory position of the two defects. $x_8$ and $x_9$ also correspond to fluctuating degrees of freedom of the bulk probe. The operators dual to these fluctuations are the real and imaginary part of the fermion bilinear operator (that is $\bar{\psi}^i \psi_j$ and
$\bar{\psi}^i \gamma_5 \psi_j$). Due to the non-Abelian structure, we can not rotate away the imaginary part anymore. Real and imaginary parts will be genuinely non-commuting. Note that once we fixed $x_3$ to be diagonal as a 2 by 2 matrix, we no longer have the freedom to diagonalize these 2 operators. We can tag fermion 1 and fermion 2 to be the matter content localized on the two spatially separated defects respectively. Diagonal mass terms give a mass to the fermion on one or the other defect. Off-diagonal mass terms however are non-local - they mix a fermion on defect 1 with a fermion on defect 2. To be gauge invariant under $SU(N_c)$ gauge transformations, such a bi-local operator needs to include a Wilson line connecting the two defects just like the chiral condensate operator in the Sakai-Sugimoto model \cite{Sakai:2004cn}.

This origin of the non-Abelian nature of our configuration is the main contrast to the standard Myers effect. We are polarizing the brane by turning on a source for the boundary Wilson-line-threaded fermion bilinear on the field theory side rather than by a flux.

To arrive at a dictionary below for $x_8$ and $x_9$, we observe that the the UV behavior changes when going from the Abelian to the non-Abelian configuration:
\beq
\begin{cases} 1/\rho\\ 1
\end{cases} \rightarrow \begin{cases} e^{-m \rho }/ \rho \quad \text{normalizable}\\
e^{m \rho }/ \rho \quad \text{non-normalizable}
\end{cases}
\eeq
The non-normalizable mode on the field theory side will dual to a source for Wilson line operator. In the bulk the off-diagonal modes of multi-brane configuration, would normally be interpreted as open string stretched between two defects. As we will see more concretely below, the exponential factor accompanying the standard power can be seen as an avatar of the string worldsheet.

\section{Wilson line two point function from two sides of Duality }

Given our matrix valued brane configuration in the bulk, we can turn on a source for the Wilson-line-threaded fermion bilinear on the field theory side and can calculate the corresponding two point function. For simplicity we will refer to this Wilson-line-threaded fermion bilinear operator from now on as a short Wilson line. The first step is to find the propagator, which is described by a differential equation in the 3 field theory direction $\overrightarrow{x}$ and the AdS radial direction $\rho$:
\begin{eqnarray}
(\square_G+2-G_{33} m^2)&&  G(\rho', \overrightarrow{x}',\rho, \overrightarrow{x}) \nonumber\\ = \delta(\rho', \overrightarrow{x}',\rho, \overrightarrow{x}) &&
\eqn{EOM}\end{eqnarray}
For generic source this is referred to as the bulk-to-bulk propagator. In the limit that the source is taken to the boundary, it is referred to the bulk-to-boundary propagator. In the latter case, the $\delta$-function source can be replaced by a boundary condition on $G$.
$\square_G$ is the scalar Laplacian for theory on the probe brane
\begin{eqnarray}
\square_G&=&\frac{1}{\sqrt{g}}\partial_\mu \sqrt{g}g^{\mu\nu} \partial _\nu \\ &=& \rho^2\partial_{\rho}^2+4\rho \partial_{\rho}+\rho^{-2} \partial^2_{\overrightarrow{x}} \nonumber
\end{eqnarray}
$G_{33} m^2= \rho^2 m^2$ is and effective position dependent mass term due to the higher order commutators in non-Abelian DBI, which would induce non-conformal deformation to the usual conformal defect.
This is the equation for the slipping mode $\theta$, rather than $x_9=\rho \theta$ from the previous section. It is $\theta$, not $x_9$, whose fluctuations obey a standard scalar wave equations and so its use will simplify the discussion in the section. The mass squared of -2 for this slipping mode comes from the fact that the volume of the $S^2$ wrapped by the D5 brane shrinks as it slips off the equator.

We were unable to find an analytical solution to this differential equation. However, we can discuss the behavior of the solution in different limits. There are two different regimes where the bulk-to-bulk correlation function behaves qualitatively different, $m \rho \gg 1$ and $m \rho \ll 1$, corresponding to whether the non-local and hence non-conformal deformation can be ignored in the equation of motion or not.

To see this in more detail, one can look at the standard solution for the bulk-to-boundary propagator without non-Abelian deformation:
\beq
(\square_G +2) G(\rho, \overrightarrow{x})= 0
\eeq
\beq
\label{abelianprop}
G(\rho, \overrightarrow{x}) = \frac{1}{\rho^2}\frac{1}{(\overrightarrow{x}^2+ 1/\rho^2)^2}
\eqn{sol}\eeq
In the second regime, $m \rho \gg 1$, one can ignore the non-Abelian deformation and so \eqref{abelianprop} is the correct bulk-to-boundary correlator. The equation of motion is just that for the standard slipping mode, dual to fermion bilinear in field theory side. In the field theory, its two point correlation function goes as $1/x_d^4$ with $x_d$ being the separation between the two short Wilson lines along the defect. This corresponds to conformal dimension $\Delta =2$ as seen above. So this $m \rho \ll 1$ regime is the Abelian regime.

The condition on the first regime, $m \rho \gg 1$, can be equivalently written as $ d \rho \gg l_s^2$, using our result for $m$ in terms of $l_s$, \eqref{mresult}. Note that this is exactly the regime in which the standard description of the Wilson line in terms of a classical string world sheet calculation following \cite{Maldacena:1998im} is valid. Including the warp factor, $\rho d$ is the effective width of world sheet and the condition that $ d \rho \gg l_s^2$ hence guarantees that the string worldsheet is classical.

\begin{figure}[t]
\includegraphics[width=6cm]{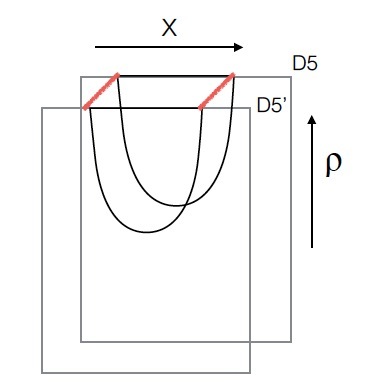}
\caption{Two thick lines are the short Wilson line, with a string world sheet hanging down and stretching between them. The world sheet ends on D-branes in the bulk }
\label{fig:Ushape}
\end{figure}

In this regime, the two point function of two spatially separated short Wilson lines can hence be captured by the area of a string world sheet connecting those two Wilson line. The world sheet, displayed in figure \ref{fig:Ushape}, ends on the two parallel branes in the bulk and connects the two Wilson lines on the boundary. From Maldacena's calculation \cite{Maldacena:1998im} of the potential of two quarks, the area of this string world sheet is proportional to $1/x_d$. It follows that for the configuration of two such Wilson lines as displayed in figure \ref{fig:Ushape} the correlation is:
\beq
\nonumber \langle W W \rangle  \sim  e^{-S} \sim e^ {\frac{C}{ x_d}}\\
C=\frac{4\pi^2 (2\lambda)^\frac{1}{2}}{\Gamma (1/4)^4}.
\eqn{WW}
\eeq

In order to confirm that our description of the system in terms of a non-Abelian brane configuration describes this physics correctly, we would like to reproduce \eq{WW} from the non-Abelian DBI. Note that the effective mass appearing in our scalar wave equation, $M = \sqrt{m^2 \rho^2 -2} \sim m \rho$, is large in this regime and so we are asked to calculate the propagator for a very massive point particle. Furthermore, as indicated, the position independent contribution of the mass can be neglected in this regime.
In the large mass limit solutions to the scalar wave equations can be reduced to the problem of finding geodesics using the WKB approximation. For a recent detailed discussion of this relation see \cite{Festuccia:2005pi}. This standard derivation also makes it clear that in the case of interest here, a scalar with a position dependent mass $M=m \rho$, we simply have to find the appropriate generalized geodesic following from an action
\begin{equation}
S_{\text{particle}} = \int M(\rho) ds = m \int \rho ds.
\end{equation}
with
\beq
ds^2=\rho^2 (dt^2+d\vec{x}^2)+\frac{d\rho^2}{\rho^2}.
\eeq
In terms of the geodesic action, the 2-pt function of the Wilson lines is given by
\beq
\nonumber \langle W W \rangle  \sim  e^{-S_{\text{particle}}}
\eeq

We can take the direction of separation along the field theory directions as $x$, as shown in figure \ref{fig:Ushape}. With this the generalized point particle action becomes:
\beq
S_{\text{particle}}= \frac{2d}{\lambda}\int_{\rho_1}^{\rho_2}  \sqrt{(\partial_x \rho)^2+\rho^4} \ dx
\eeq
$2d$ being the length of the short Wilson line or, equivalently, the width of the world sheet. But this is exactly the action for the string world sheet sourced by two parallel moving quarks as in \cite{Maldacena:1998im}! For the string worldsheet, the determinant of the induced metric had an extra power of $\rho$ compared to the standard geodesic simply because the worldsheet is 2-dimensional and so picked up an extra warpfactor. In the geodesic action the same factor of $\rho$ appears due to the position dependent mass that we inherited from the leading commutator term in the non-Abelian DBI.

In the bulk we saw that the two regimes, Abelian versus WKB, are separated by
\beq
m\rho \sim 1.
\eeq
The correlation function of the short Wilson loop goes as $x_d^{-4}$ and with the exponential of \eq{WW} respectively in those two regimes, and presumably smoothly interpolates between these two behaviors in between. We would like to know for what separation $x_d$ these two calculations are valid.
Looking at the solution for the worldsheet, one can see that the turn around point, $\rho_*$, and the field theory separation satisfy the relation $x_d \rho_* \sim 1$. This implies $m \rho_*  \sim  \frac{d}{x l_s^2}$. The two behaviors should cross over into each other when $m \rho_* \sim 1$.
This implies that on the field theory side, the two behaviors are separated by $x_* \sim d \sqrt{\lambda_t}$. Here we used that in our $R=1$ units we have $l_s^{-2} = \sqrt{\lambda_t}$, the 't Hooft coupling, according to the standard AdS/CFT dictionary. For $x_d \gg x_*$ we see that standard $x_d^{-4}$ fall-off of the two-point function, for $x_d \ll x_*$ the exponential behavior of \eq{WW}.

\section{Non-supersymmetric D3/D5 with axion flux}

We can also consider a slightly different set-up, where the non-Abelian features implied by the non-local mass terms we introduced above are enhanced by coupling to a 9-form flux via the Chern-Simon term. This time, take the $N_f$ D5-branes to intersect the $N_c$ D3-branes in $1+1$ dimensions. In addition, a spatially varying axion field is turned on corresponding to a constant 1-form field-strength. The dual 9-form flux can be thought of as being sourced by smeared D7 branes behind the horizon \cite{Azeyanagi:2009pr}. This set-up is summarized in the following table:
\begin{eqnarray}
\begin{array}{c|rrrrrrrrrr|}
  {}  & 0  &1 &2 &3 &4& 5  & 6 & 7 & 8& 9 \\ \hline
 D3 & \times& \times& \times& \times&&&&&&\\
  D5 & \times& \times&&& \times&\times& \times& \times& &\\\hline
F9 & \times& \times& \times& & \times& \times& \times& \times& \times& \times\\
\hline
\end{array}\end{eqnarray}
The gravity description will be once more be studied in the probe limit, with $N_f\ll N_c$.

The interpretation of the flux is best understood when writing the flux as a 1-form flux, that is a constant gradient of the string theory axion. In the solution of \cite{Azeyanagi:2009pr} the axion is linear in $x_3$ and independent of all other coordinates. This constant 1-form flux is dual to a position dependent theta angle in the field theory, where theta grows linearly with $x_3$. We take all our defects to live at the same $x_3$, so they all feel the same theta angle. The non-trivial gradient in theta still has some non-trivial effect on the defects. To impose the non-Abelian structure we separate our defects along the $x_2$ direction and once more turn on non-Abelian masses.

\subsection{Single brane solution in pure AdS}
Let us first consider the AdS/CFT dual of our second D3-D5 brane system with a single D5 brane probe and ignore 9-flux for now.
In this case, the background geometry is still the asymptotically flat, zero temperature D3-brane solution \eqref{d3metric}.
This time we rewrite $\Omega_5$ as:
\beq
dr^2+r^2 d\Omega_5^2=d^2\rho+\rho^2 d\Omega_3^2+d^2x_8+d^2x_9
\eeq
with
\beq
\eqn{fis} f=\frac{R^4}{(\rho^2+x_8^2+x_9^2)^2} = \frac{R^4}{r^4}.
\eeq
In what follows we will again work in units where $R=1$.
The D5 probe brane spans $x_0$, $x_1$ and wraps the $S^3$ above and lives on a curve $x_9(\rho)$. For the simplest solution we set $x_5=x_3=0$. The pull-back metric on the D5 brane worldvolume is:
\beq
ds_6^2=\frac{(-dt^2+dx_1^2)}{f^{1/2}}+f^{\frac{1}{2}} \left [ (1+x_9'^2)d\rho^2+ \rho^2 d\Omega_3^2
\right ] \eeq
Without worldvolume gauge field, Born-Infeld action is:
\begin{eqnarray}
S_{D5}&=&-N_fT_5\int d^6\xi \sqrt{-g_6}
\\ \nonumber &=&-N_fT_5 Vol(S^3)\int d^2x d\rho f^{\frac{1}{2}}\rho^3\sqrt{1+x_9'^2}
\end{eqnarray}
$x_9=0$ is the ground state solution describing a D5 brane probe on AdS$_3$ $\times$ S$^3$.
From the linearized equations of motion for $x_9$ one can read off its asymptotic behavior:
\beq
x_9(\rho)=C_1 \rho^{\sqrt{2}i}+C_2\rho^{-\sqrt{2}i}.
\eeq
As in the supersymmetric brane configuration, this fluctuation is dual to the fermion mass term on the defect. The fact that the exponents are imaginary tells
us that while this operator has dimension 1 in the free field theory, at strong coupling it has naively an imaginary dimension in contradiction with unitarity. Such an asymptotic behavior is due to the fact that the slipping mode with a mass squared of $-3$ in AdS unit is below the AdS$_3$ BF bound  of -1 and so really signals an instability of the system, not a loss of unitarity.
The analysis of our AdS$_3$ $\times$ $S^3$ brane is very similar to the quantum hall plateau transition studied in \cite{Davis:2008nv}. There also the system is unstable in pure AdS. In \cite{Davis:2008nv} the asymptotic behavior was tamed by replacing AdS with the full D3 brane metric, that is the function $f$ above was replaced with  $f=1+\frac{R^4}{(\rho^2+x_9^2+x_5^2)^2}$. To truly stabilize the system it was later realized that internal fluxes can be turned on \cite{Myers:2008me,Bergman:2010gm}. This should also be possible here.

\subsection{Interpolating Solution between AdS5 and D3-D7 Scaling Solution in Type-IIB supergravity}

In this section we would like to review some of the crucial aspects of the type IIB supergravity solution with the fully backreacted 9-form flux following mostly the original presentation in \cite{Azeyanagi:2009pr}. The whole solution is driven by a constant axion gradient
\beq
F_1 = d \chi = \beta dx_3
\eeq
together with the standard D3-brane flux already present in AdS$_5$ $\times$ $S^5$.
In the field theory this extra 1-form flux describes a constant gradient of the theta angle, $\theta=\beta c_3$. As the theta angle is dimensionless, $\beta$ carries the dimension of energy - it sets a scale in the problem.
To construct the full solution one parametrizes the (Einstein frame) metric as:
\begin{eqnarray}
\label{flow}
ds^2_{10}&=&f^{\frac{1}{2}}e^{2b(r)-2c(r)-\frac{\phi_0}{2}}(-dt^2+d\vec{x}^2)+ \\ \nonumber && f^{\frac{3}{2}}e^{2b(r)-6c(r)-\frac{3\phi_0}{2}}dx_3^2
+ \\ \nonumber && e^{-\frac{\phi_0}{2}}f^{\frac{1}{2}}(d^2\rho+\rho^2d\Omega_3^2+d^2x_5+d^2x_9)
\end{eqnarray}
with $f$ and $r$ still given by \eq{fis}. The dilaton is parametrized by
\beq \phi(r)=4c(r)+4 \log r+\phi_0 .\eeq
For
\beq
b(r) = \log r, \quad c(r)= - \log r, \quad \phi_0=0
\eeq
this becomes the standard AdS$_5$ $\times$ $S^5$ metric of \eq{G}. At was shown in
\cite{Azeyanagi:2009pr} a full solution to Einstein's equations with the constant axion gradient interpolates
between this AdS$_5$ $\times$ $S^5$ solution in the UV (at large $\rho$) to an IR ``scaling solution" in which
$b \sim 7 (\log r)/\sqrt{33}$ and $c \sim (\log r)/\sqrt{33}$. That is the solution describes a flow from ${\cal N}=4$ SYM in the UV, which means energies much larger than $\beta$, to a scale invariant, gapless theory in the IR with
an anisotropic scaling exponent ($x_3$ scales differently from $t$, $x_1$ and $x_2$). The full interpolating flow for $b(r)$ and $c(r)$ can easily be generated numerically, as described in detail in \cite{Azeyanagi:2009pr}. We reproduced this numerical solution and used it as a background to embed our probe branes in.

\subsection{Non-Abelian double brane solution}

Having reviewed all the ingredients that go into the construction, we can now present our solution for the non-Abelian embedding for multiple D5 branes.
To describe multiple coincident D5-branes we once more need the non-Abelian generalization of the Born-Infeld action \eqref{action}. The DBI part of the action is still given by
\begin{eqnarray}
S_{D5}&=&-T_5 \int d^6x  \label{action2}\\  \nonumber && Tr \left ( e^{\phi(x_i)/2} \sqrt{-det(P[G_{ab}])det(Q^i_j)} \right )
\end{eqnarray}
with $Q^i_j$ still being given by \eqref{qis}.
We this time included the coupling to the dilaton, which is no longer constant in the geometry of \eqref{flow}.
The standard $e^{-\phi}$ prefactor from the string brane DBI becomes $e^{\phi/2}$ in the Einstein frame.
Also note that the dilaton factor now has $x_i$ dependence from the back-reaction of the flux.

Taking $x_3$, $x_8$, and $x_9$ as function of $\rho$, the induced metric on the D5-brane is:
\begin{eqnarray}
ds^2_6 &=&f^{\frac{1}{2}}e^{2b(r)-2c(r)-\frac{\phi_0}{2}}(-dt^2+d^2x_1)\\ \nonumber
&& +f^{\frac{1}{2}}e^{-\frac{\phi_0}{2}}(1+x'^2_5+x'^2_9+e^{2b(r)-2c(r)}x_2'^2)d^2\rho\\
\nonumber &&+f^{\frac{1}{2}}\rho^2d\Omega_3^2
\end{eqnarray}
and
the Abelian part of the action \eqref{action2} yields:
\begin{eqnarray}
S_{D5}&=&-T_5 \int d\Omega_3 d^2xd\rho  e^{\frac{\phi(r)}{2}}Tr\sqrt{-g_6} \nonumber \\
&=&-T_5\int d\Omega_3 d^2xd\rho \, \rho^3 f^{3/2} e^{2b} r^2 \label{abelian} \\ &&
 \nonumber  Tr\sqrt{1+x'^2_5+x'^2_9+e^{2b(r)-2c(r)}x_2'^2}.
\end{eqnarray}

The polarization of the probe brane is aided by the background RR field. The RR field couples directly to the $N_f$ D-branes via a Chern-Simons term.
\beq
S_{CS} = \frac{e^{2\phi}}{2 \pi \alpha'}\frac{i}{3}\mu_5\int dt Tr(x^i x^j x^k)F^{(9)}_{tijk\mu_1 \ldots \mu_5 }
\eeq
Here, and in what follows, $i$, $j$, and $k$ will run over indices 2, 5, and 9, whereas $\mu_1, \ldots \mu_5$ run over the 5 worldvolume directions.
The indices of the 9-form field strength $F^{(9)}$ in the linear axion solution run over $x_0 \sim x_9$ without the $x_3$ direction. This flux promotes a polarization of the D5 brane coordinates $x_2$, $x_5$, and $x_9$ into matrix valued coordinates. As we mentioned before, commutator terms from the DBI action only arise at quartic order in the fields. The net-effect of the 9-form flux is to introduce the non-commutativity already at cubic order in the action. In the limit of small fields, this makes the non-commutativity more pronounced.

With $F^{(9)}_{tijk \mu_1 \ldots \mu_9}$ coming from the constant 1-form from the last section it is, in components proportional to the spacetime metric $g^{33}\sqrt{-g}=\rho^3 f^{3/2}e^{2b(r)}$:
\beq
F^{(9)}_{tijk \mu_1 \ldots \mu_5}=
  \beta \rho^3 f^{3/2}e^{2b(r)}\epsilon_{ijk} \epsilon_{\mu_1 \ldots \mu_5}
\eeq

As for the supersymmetric defect, we can only reliably study the non-Abelian action by keeping the first few terms of a power series expansion in small fields. As we reviewed for the supersymmetric defect, the DBI part of the action has the first commutators showing up at quartic order in the fields. The background flux introduces a cubic commutator term in the CS terms. So expanding the action to cubic order in the fields will allow us to confine the non-commutative nature of the solution to the CS term and use the Abelian action \eqref{abelian} for the DBI part. The only non-trivial commutator in the equations of motion following from the cubic action comes from the contribution
of the CS term:
\beq
\ldots +ie^{2\phi(r)}\rho^3\beta  f^{3/2}e^{2b(r)}\epsilon_{ijk}[x_j,x_k]=0
\eeq
where the $\ldots$ stand for the terms from the DBI part of the action.

Let us consider the same $SU(2)$ ansatz for the non-commuting solution we used before: $x_2=X_{1}(\rho)\sigma_3$, $x_8=X_{2}(\rho)\sigma_1$, $x_9=X_{3}(\rho)\sigma_2$, with $X_2=X_3$.
The background spacetime is asymptotic AdS in the UV, so we can use the metric \eq{G} for the asymptotic analysis and work with constant dilaton. We again take an ansatz $X_{1}(\rho)\rightarrow d+\epsilon g(\rho), X_{2,3}(\rho)\rightarrow\epsilon k(\rho)$, so that the two defects are sitting at $x_2 = \pm d$ respectively. For this solution to be captured by a small field expansion, we need to work with sufficiently small $d$. Expanding to first order in $\epsilon$, one arrives at two equations:
\beq
\nonumber &-d \beta p k(\rho) + K [2p k(\rho)+ p^2 (k'(\rho)+\rho k''(\rho))]=0\\
&5g'(\rho)+\rho g''(\rho)=0
\eeq
which yield solutions:
\beq
&g(\rho)=\frac{C_1}{4\rho^4}\\
&k(\rho)=C'_1 \rho^{\sqrt{-2+\frac{d}{K} \beta}}+C'_2 \rho^{-\sqrt{-2+\frac{d}{K} \beta}}
\eeq
The behavior of $g$ is as in the Abelian case.

With
\beq -2+\frac{d}{K} \beta = -2 + \frac{m}{3}  \beta >0 \eqn{dimension} \eeq
the equations have stable solutions. Here
\beq
m  = \frac{2 d}{2 \pi \alpha'}
\eeq
is again the mass associated with a string stretched across the distance $2d$ setting the separation of the two defects. Interestingly, the separation allows us to stabilize the slipping mode for small brane separation as long as the flux is large. Of course, for sufficiently large brane separation the slipping mode will always be stabilized by $d^2$ terms arising from the quartic commutator squared terms in the DBI action. But this would be outside the regime where we understand the DBI.

To find solutions that smoothly cap off in the infrared we need to consider an action where higher powers of the derivatives are non-negligible and we need to go beyond the rigorously understood aspects of the DBI. An ad-hoc ansatz we will employ for the action is to keep the full non-linear structure under the square root
\beq
\eqn{fulldbi}
{\cal L}_{DBI}=-{\cal F} \sqrt{(1+x'^2_5+x'^2_9+e^{2b(r)-2c(r)}x_2'^2)}
\eeq
but to neglect all commutator terms. This prescription seems to essentially agree with the adapted symmetrized trace prescription that was used in \cite{Erdmenger:2007ja} to find p-wave superconductors using non-Abelian gauge fields on probe branes.
Note that, in either case, with all $\sigma_i^2=1$, the matrix structure match between DBI and WZ term and can be scaled out from the equations of motion.
We seek a solution that smoothly truncates at a finite value of the radial coordinate (a gapped or ``Minkowski" embedding). Even with the ad-hoc action, finding the full brane embedding can only be done numerically. This is technically challenging since the to be solved for functions $x_8(\rho)$ and $x_9(\rho)$ appear as arguments of the numerically obtained metric functions $b(r)$ and $c(r)$. To obtain a Minkowski embedding we start $x_2$ with very large derivatives (ideally infinite) at a finite value $\rho_*$. For an Abelian embedding this would follow from the requirement that the two separate branes dual to the two separated defects at $x_2 = \pm d$ smoothly connect into a single connected configuration $x_2(\rho)$ that close to the endpoint is given by a square root. $x_{8,9}$ on the other hand start with a finite value at $\rho_*$, but we require
zero derivative. In the Abelian case this would correspond to the requirement that the two branches of the solution (with positive and negative $x_2$ respectively) are symmetric under exchange. One solution is shown in Figure \ref{fig:fg}:
\begin{figure}[h]
\includegraphics[width=7cm]{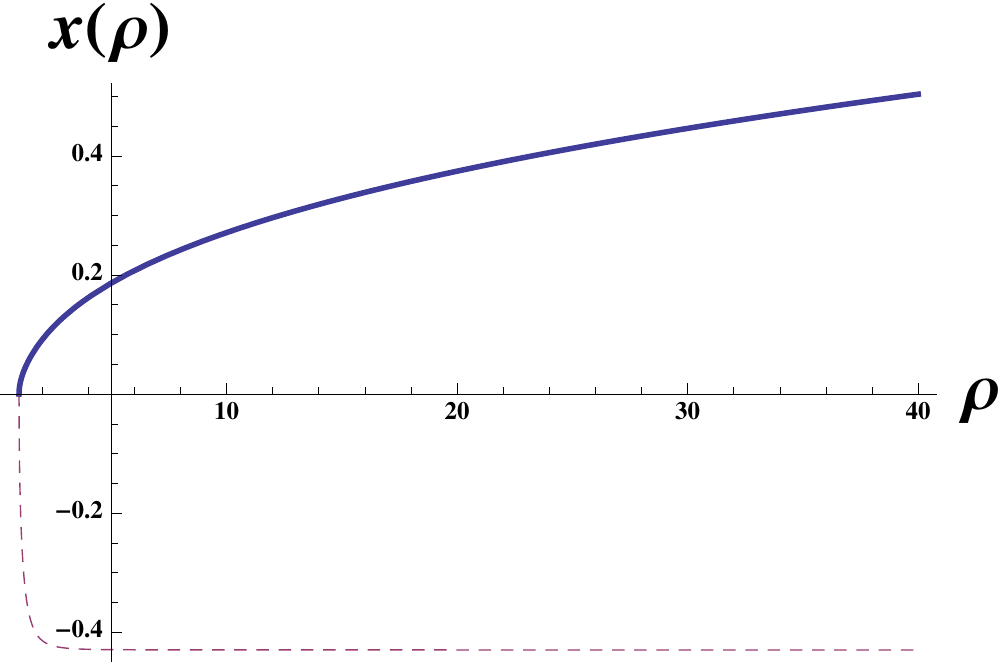}
\caption{\label{fig:profile} {Solid line represents $x_{8,9}$ function, while dashed is $x_2$. For $\frac{\beta}{K}=-50$ and $d=-4.3\times10^{-2}$, Shooting out solution at $\rho_*=1$, $X_5(\rho_*)=X_{8,9}(\rho_*)=0$}}
\label{fig:fg}
\end{figure}
We used $\beta/K=-50$ to generate this plot giving rise to $d=-4.3\times10^{-2}$ in the asymptotic solution.
The asymptotic form of the solution in the UV agrees with our analytic answer \eq{dimension}.
This solution mostly serves as a proof of principle.

\section{Conclusion}
We have shown that certain non-local mass terms in a field theory with an holographic dual in terms of probe branes can force the geometry of the probe to be non-geometric. The embedding coordinates are genuinely matrix valued. We demonstrated this basic effect in a supersymmetric defect setup, where we were also able to calculate the short-distance behavior of a two-point function using the non-Abelian DBI action. We also demonstrated that, if we turn on in addition a brane polarizing background flux, the effect is enhanced as expected from Myers' proposal.

\section*{Acknowledgments}
We would like to thank Joe Polchinski for many helpful discussions and suggestions, and Chris Herzog for a suggestion on PDE. This work is supported, in part, by the US Department of Energy under grant number DE-FG02-96ER40956. S.S. was supported in part by the National Science Foundation under Grant No. PHY11-25915 and gratefully acknowledges a graduate fellowship at the Kavli Institute for Theoretical Physics.

\bibliography{Brbrane} 

\begin{thebibliography}{19}%
\makeatletter
\providecommand \@ifxundefined [1]{%
 \@ifx{#1\undefined}
}%
\providecommand \@ifnum [1]{%
 \ifnum #1\expandafter \@firstoftwo
 \else \expandafter \@secondoftwo
 \fi
}%
\providecommand \@ifx [1]{%
 \ifx #1\expandafter \@firstoftwo
 \else \expandafter \@secondoftwo
 \fi
}%
\providecommand \natexlab [1]{#1}%
\providecommand \enquote  [1]{``#1''}%
\providecommand \bibnamefont  [1]{#1}%
\providecommand \bibfnamefont [1]{#1}%
\providecommand \citenamefont [1]{#1}%
\providecommand \href@noop [0]{\@secondoftwo}%
\providecommand \href [0]{\begingroup \@sanitize@url \@href}%
\providecommand \@href[1]{\@@startlink{#1}\@@href}%
\providecommand \@@href[1]{\endgroup#1\@@endlink}%
\providecommand \@sanitize@url [0]{\catcode `\\12\catcode `\$12\catcode
  `\&12\catcode `\#12\catcode `\^12\catcode `\_12\catcode `\%12\relax}%
\providecommand \@@startlink[1]{}%
\providecommand \@@endlink[0]{}%
\providecommand \url  [0]{\begingroup\@sanitize@url \@url }%
\providecommand \@url [1]{\endgroup\@href {#1}{\urlprefix }}%
\providecommand \urlprefix  [0]{URL }%
\providecommand \Eprint [0]{\href }%
\providecommand \doibase [0]{http://dx.doi.org/}%
\providecommand \selectlanguage [0]{\@gobble}%
\providecommand \bibinfo  [0]{\@secondoftwo}%
\providecommand \bibfield  [0]{\@secondoftwo}%
\providecommand \translation [1]{[#1]}%
\providecommand \BibitemOpen [0]{}%
\providecommand \bibitemStop [0]{}%
\providecommand \bibitemNoStop [0]{.\EOS\space}%
\providecommand \EOS [0]{\spacefactor3000\relax}%
\providecommand \BibitemShut  [1]{\csname bibitem#1\endcsname}%
\let\auto@bib@innerbib\@empty
\bibitem [{\citenamefont {Myers}(1999)}]{Myers:1999ps}%
  \BibitemOpen
  \bibfield  {author} {\bibinfo {author} {\bibfnamefont {R.~C.}\ \bibnamefont
  {Myers}},\ }\href@noop {} {\bibfield  {journal} {\bibinfo  {journal} {JHEP}\
  }\textbf {\bibinfo {volume} {9912}},\ \bibinfo {pages} {022} (\bibinfo {year}
  {1999})},\ \Eprint {http://arxiv.org/abs/hep-th/9910053}
  {arXiv:hep-th/9910053 [hep-th]} \BibitemShut {NoStop}%
\bibitem [{\citenamefont {Maldacena}(1998{\natexlab{a}})}]{Maldacena:1997re}%
  \BibitemOpen
  \bibfield  {author} {\bibinfo {author} {\bibfnamefont {J.~M.}\ \bibnamefont
  {Maldacena}},\ }\href {\doibase 10.1023/A:1026654312961,
  10.1023/A:1026654312961} {\bibfield  {journal} {\bibinfo  {journal}
  {Adv.Theor.Math.Phys.}\ }\textbf {\bibinfo {volume} {2}},\ \bibinfo {pages}
  {231} (\bibinfo {year} {1998}{\natexlab{a}})},\ \Eprint
  {http://arxiv.org/abs/hep-th/9711200} {arXiv:hep-th/9711200 [hep-th]}
  \BibitemShut {NoStop}%
\bibitem [{\citenamefont {Gubser}\ \emph {et~al.}(1998)\citenamefont {Gubser},
  \citenamefont {Klebanov},\ and\ \citenamefont {Polyakov}}]{Gubser:1998bc}%
  \BibitemOpen
  \bibfield  {author} {\bibinfo {author} {\bibfnamefont {S.}~\bibnamefont
  {Gubser}}, \bibinfo {author} {\bibfnamefont {I.~R.}\ \bibnamefont
  {Klebanov}}, \ and\ \bibinfo {author} {\bibfnamefont {A.~M.}\ \bibnamefont
  {Polyakov}},\ }\href {\doibase 10.1016/S0370-2693(98)00377-3} {\bibfield
  {journal} {\bibinfo  {journal} {Phys.Lett.}\ }\textbf {\bibinfo {volume}
  {B428}},\ \bibinfo {pages} {105} (\bibinfo {year} {1998})},\ \Eprint
  {http://arxiv.org/abs/hep-th/9802109} {arXiv:hep-th/9802109 [hep-th]}
  \BibitemShut {NoStop}%
\bibitem [{\citenamefont {Witten}(1998)}]{Witten:1998qj}%
  \BibitemOpen
  \bibfield  {author} {\bibinfo {author} {\bibfnamefont {E.}~\bibnamefont
  {Witten}},\ }\href@noop {} {\bibfield  {journal} {\bibinfo  {journal}
  {Adv.Theor.Math.Phys.}\ }\textbf {\bibinfo {volume} {2}},\ \bibinfo {pages}
  {253} (\bibinfo {year} {1998})},\ \Eprint
  {http://arxiv.org/abs/hep-th/9802150} {arXiv:hep-th/9802150 [hep-th]}
  \BibitemShut {NoStop}%
\bibitem [{\citenamefont {Karch}\ and\ \citenamefont
  {Randall}(2001)}]{Karch:2000gx}%
  \BibitemOpen
  \bibfield  {author} {\bibinfo {author} {\bibfnamefont {A.}~\bibnamefont
  {Karch}}\ and\ \bibinfo {author} {\bibfnamefont {L.}~\bibnamefont
  {Randall}},\ }\href@noop {} {\bibfield  {journal} {\bibinfo  {journal}
  {JHEP}\ }\textbf {\bibinfo {volume} {0106}},\ \bibinfo {pages} {063}
  (\bibinfo {year} {2001})},\ \Eprint {http://arxiv.org/abs/hep-th/0105132}
  {arXiv:hep-th/0105132 [hep-th]} \BibitemShut {NoStop}%
\bibitem [{\citenamefont {Karch}\ and\ \citenamefont
  {Katz}(2002)}]{Karch:2002sh}%
  \BibitemOpen
  \bibfield  {author} {\bibinfo {author} {\bibfnamefont {A.}~\bibnamefont
  {Karch}}\ and\ \bibinfo {author} {\bibfnamefont {E.}~\bibnamefont {Katz}},\
  }\href@noop {} {\bibfield  {journal} {\bibinfo  {journal} {JHEP}\ }\textbf
  {\bibinfo {volume} {0206}},\ \bibinfo {pages} {043} (\bibinfo {year}
  {2002})},\ \Eprint {http://arxiv.org/abs/hep-th/0205236}
  {arXiv:hep-th/0205236 [hep-th]} \BibitemShut {NoStop}%
\bibitem [{\citenamefont {Sakai}\ and\ \citenamefont
  {Sugimoto}(2005)}]{Sakai:2004cn}%
  \BibitemOpen
  \bibfield  {author} {\bibinfo {author} {\bibfnamefont {T.}~\bibnamefont
  {Sakai}}\ and\ \bibinfo {author} {\bibfnamefont {S.}~\bibnamefont
  {Sugimoto}},\ }\href {\doibase 10.1143/PTP.113.843} {\bibfield  {journal}
  {\bibinfo  {journal} {Prog.Theor.Phys.}\ }\textbf {\bibinfo {volume} {113}},\
  \bibinfo {pages} {843} (\bibinfo {year} {2005})},\ \Eprint
  {http://arxiv.org/abs/hep-th/0412141} {arXiv:hep-th/0412141 [hep-th]}
  \BibitemShut {NoStop}%
\bibitem [{\citenamefont {Maldacena}(1998{\natexlab{b}})}]{Maldacena:1998im}%
  \BibitemOpen
  \bibfield  {author} {\bibinfo {author} {\bibfnamefont {J.~M.}\ \bibnamefont
  {Maldacena}},\ }\href {\doibase 10.1103/PhysRevLett.80.4859} {\bibfield
  {journal} {\bibinfo  {journal} {Phys.Rev.Lett.}\ }\textbf {\bibinfo {volume}
  {80}},\ \bibinfo {pages} {4859} (\bibinfo {year} {1998}{\natexlab{b}})},\
  \Eprint {http://arxiv.org/abs/hep-th/9803002} {arXiv:hep-th/9803002 [hep-th]}
  \BibitemShut {NoStop}%
\bibitem [{\citenamefont {Polchinski}\ and\ \citenamefont
  {Strassler}(2002)}]{Polchinski:2001tt}%
  \BibitemOpen
  \bibfield  {author} {\bibinfo {author} {\bibfnamefont {J.}~\bibnamefont
  {Polchinski}}\ and\ \bibinfo {author} {\bibfnamefont {M.~J.}\ \bibnamefont
  {Strassler}},\ }\href {\doibase 10.1103/PhysRevLett.88.031601} {\bibfield
  {journal} {\bibinfo  {journal} {Phys.Rev.Lett.}\ }\textbf {\bibinfo {volume}
  {88}},\ \bibinfo {pages} {031601} (\bibinfo {year} {2002})},\ \Eprint
  {http://arxiv.org/abs/hep-th/0109174} {arXiv:hep-th/0109174 [hep-th]}
  \BibitemShut {NoStop}%
\bibitem [{\citenamefont {Kristjansen}\ and\ \citenamefont
  {Semenoff}(2012)}]{Kristjansen:2012ny}%
  \BibitemOpen
  \bibfield  {author} {\bibinfo {author} {\bibfnamefont {C.}~\bibnamefont
  {Kristjansen}}\ and\ \bibinfo {author} {\bibfnamefont {G.~W.}\ \bibnamefont
  {Semenoff}},\ }\href@noop {} {\  (\bibinfo {year} {2012})},\ \Eprint
  {http://arxiv.org/abs/1212.5609} {arXiv:1212.5609 [hep-th]} \BibitemShut
  {NoStop}%
\bibitem [{\citenamefont {Erdmenger}\ \emph {et~al.}(2008)\citenamefont
  {Erdmenger}, \citenamefont {Kaminski},\ and\ \citenamefont
  {Rust}}]{Erdmenger:2007ja}%
  \BibitemOpen
  \bibfield  {author} {\bibinfo {author} {\bibfnamefont {J.}~\bibnamefont
  {Erdmenger}}, \bibinfo {author} {\bibfnamefont {M.}~\bibnamefont {Kaminski}},
  \ and\ \bibinfo {author} {\bibfnamefont {F.}~\bibnamefont {Rust}},\ }\href
  {\doibase 10.1103/PhysRevD.77.046005} {\bibfield  {journal} {\bibinfo
  {journal} {Phys.Rev.}\ }\textbf {\bibinfo {volume} {D77}},\ \bibinfo {pages}
  {046005} (\bibinfo {year} {2008})},\ \Eprint {http://arxiv.org/abs/0710.0334}
  {arXiv:0710.0334 [hep-th]} \BibitemShut {NoStop}%
\bibitem [{\citenamefont {Myers}\ and\ \citenamefont
  {Wapler}(2008)}]{Myers:2008me}%
  \BibitemOpen
  \bibfield  {author} {\bibinfo {author} {\bibfnamefont {R.~C.}\ \bibnamefont
  {Myers}}\ and\ \bibinfo {author} {\bibfnamefont {M.~C.}\ \bibnamefont
  {Wapler}},\ }\href {\doibase 10.1088/1126-6708/2008/12/115} {\bibfield
  {journal} {\bibinfo  {journal} {JHEP}\ }\textbf {\bibinfo {volume} {0812}},\
  \bibinfo {pages} {115} (\bibinfo {year} {2008})},\ \Eprint
  {http://arxiv.org/abs/0811.0480} {arXiv:0811.0480 [hep-th]} \BibitemShut
  {NoStop}%
\bibitem [{\citenamefont {Ammon}\ \emph {et~al.}(2009)\citenamefont {Ammon},
  \citenamefont {Erdmenger}, \citenamefont {Kaminski},\ and\ \citenamefont
  {Kerner}}]{Ammon:2009fe}%
  \BibitemOpen
  \bibfield  {author} {\bibinfo {author} {\bibfnamefont {M.}~\bibnamefont
  {Ammon}}, \bibinfo {author} {\bibfnamefont {J.}~\bibnamefont {Erdmenger}},
  \bibinfo {author} {\bibfnamefont {M.}~\bibnamefont {Kaminski}}, \ and\
  \bibinfo {author} {\bibfnamefont {P.}~\bibnamefont {Kerner}},\ }\href
  {\doibase 10.1088/1126-6708/2009/10/067} {\bibfield  {journal} {\bibinfo
  {journal} {JHEP}\ }\textbf {\bibinfo {volume} {0910}},\ \bibinfo {pages}
  {067} (\bibinfo {year} {2009})},\ \Eprint {http://arxiv.org/abs/0903.1864}
  {arXiv:0903.1864 [hep-th]} \BibitemShut {NoStop}%
\bibitem [{\citenamefont {Azeyanagi}\ \emph {et~al.}(2009)\citenamefont
  {Azeyanagi}, \citenamefont {Li},\ and\ \citenamefont
  {Takayanagi}}]{Azeyanagi:2009pr}%
  \BibitemOpen
  \bibfield  {author} {\bibinfo {author} {\bibfnamefont {T.}~\bibnamefont
  {Azeyanagi}}, \bibinfo {author} {\bibfnamefont {W.}~\bibnamefont {Li}}, \
  and\ \bibinfo {author} {\bibfnamefont {T.}~\bibnamefont {Takayanagi}},\
  }\href {\doibase 10.1088/1126-6708/2009/06/084} {\bibfield  {journal}
  {\bibinfo  {journal} {JHEP}\ }\textbf {\bibinfo {volume} {0906}},\ \bibinfo
  {pages} {084} (\bibinfo {year} {2009})},\ \Eprint
  {http://arxiv.org/abs/0905.0688} {arXiv:0905.0688 [hep-th]} \BibitemShut
  {NoStop}%
\bibitem [{\citenamefont {Mateos}\ and\ \citenamefont
  {Trancanelli}(2011)}]{Mateos:2011ix}%
  \BibitemOpen
  \bibfield  {author} {\bibinfo {author} {\bibfnamefont {D.}~\bibnamefont
  {Mateos}}\ and\ \bibinfo {author} {\bibfnamefont {D.}~\bibnamefont
  {Trancanelli}},\ }\href {\doibase 10.1103/PhysRevLett.107.101601} {\bibfield
  {journal} {\bibinfo  {journal} {Phys.Rev.Lett.}\ }\textbf {\bibinfo {volume}
  {107}},\ \bibinfo {pages} {101601} (\bibinfo {year} {2011})},\ \Eprint
  {http://arxiv.org/abs/1105.3472} {arXiv:1105.3472 [hep-th]} \BibitemShut
  {NoStop}%
\bibitem [{\citenamefont {Bergman}\ \emph {et~al.}(2010)\citenamefont
  {Bergman}, \citenamefont {Jokela}, \citenamefont {Lifschytz},\ and\
  \citenamefont {Lippert}}]{Bergman:2010gm}%
  \BibitemOpen
  \bibfield  {author} {\bibinfo {author} {\bibfnamefont {O.}~\bibnamefont
  {Bergman}}, \bibinfo {author} {\bibfnamefont {N.}~\bibnamefont {Jokela}},
  \bibinfo {author} {\bibfnamefont {G.}~\bibnamefont {Lifschytz}}, \ and\
  \bibinfo {author} {\bibfnamefont {M.}~\bibnamefont {Lippert}},\ }\href
  {\doibase 10.1007/JHEP10(2010)063} {\bibfield  {journal} {\bibinfo  {journal}
  {JHEP}\ }\textbf {\bibinfo {volume} {1010}},\ \bibinfo {pages} {063}
  (\bibinfo {year} {2010})},\ \Eprint {http://arxiv.org/abs/1003.4965}
  {arXiv:1003.4965 [hep-th]} \BibitemShut {NoStop}%
\bibitem [{\citenamefont {DeWolfe}\ \emph {et~al.}(2002)\citenamefont
  {DeWolfe}, \citenamefont {Freedman},\ and\ \citenamefont
  {Ooguri}}]{DeWolfe:2001pq}%
  \BibitemOpen
  \bibfield  {author} {\bibinfo {author} {\bibfnamefont {O.}~\bibnamefont
  {DeWolfe}}, \bibinfo {author} {\bibfnamefont {D.~Z.}\ \bibnamefont
  {Freedman}}, \ and\ \bibinfo {author} {\bibfnamefont {H.}~\bibnamefont
  {Ooguri}},\ }\href {\doibase 10.1103/PhysRevD.66.025009} {\bibfield
  {journal} {\bibinfo  {journal} {Phys.Rev.}\ }\textbf {\bibinfo {volume}
  {D66}},\ \bibinfo {pages} {025009} (\bibinfo {year} {2002})},\ \Eprint
  {http://arxiv.org/abs/hep-th/0111135} {arXiv:hep-th/0111135 [hep-th]}
  \BibitemShut {NoStop}%
\bibitem [{\citenamefont {Festuccia}\ and\ \citenamefont
  {Liu}(2006)}]{Festuccia:2005pi}%
  \BibitemOpen
  \bibfield  {author} {\bibinfo {author} {\bibfnamefont {G.}~\bibnamefont
  {Festuccia}}\ and\ \bibinfo {author} {\bibfnamefont {H.}~\bibnamefont
  {Liu}},\ }\href {\doibase 10.1088/1126-6708/2006/04/044} {\bibfield
  {journal} {\bibinfo  {journal} {JHEP}\ }\textbf {\bibinfo {volume} {0604}},\
  \bibinfo {pages} {044} (\bibinfo {year} {2006})},\ \Eprint
  {http://arxiv.org/abs/hep-th/0506202} {arXiv:hep-th/0506202 [hep-th]}
  \BibitemShut {NoStop}%
\bibitem [{\citenamefont {Davis}\ \emph {et~al.}(2008)\citenamefont {Davis},
  \citenamefont {Kraus},\ and\ \citenamefont {Shah}}]{Davis:2008nv}%
  \BibitemOpen
  \bibfield  {author} {\bibinfo {author} {\bibfnamefont {J.~L.}\ \bibnamefont
  {Davis}}, \bibinfo {author} {\bibfnamefont {P.}~\bibnamefont {Kraus}}, \ and\
  \bibinfo {author} {\bibfnamefont {A.}~\bibnamefont {Shah}},\ }\href {\doibase
  10.1088/1126-6708/2008/11/020} {\bibfield  {journal} {\bibinfo  {journal}
  {JHEP}\ }\textbf {\bibinfo {volume} {0811}},\ \bibinfo {pages} {020}
  (\bibinfo {year} {2008})},\ \Eprint {http://arxiv.org/abs/0809.1876}
  {arXiv:0809.1876 [hep-th]} \BibitemShut {NoStop}%
\end{thebibliography}%

\end{document}